\documentclass[a4paper,12pt]{amsart}
\usepackage{amssymb,amsmath,mathrsfs}
\usepackage[usenames,dvipsnames]{xcolor}
\usepackage{hyperref}
\usepackage{tensor}
\usepackage{graphicx}
\usepackage{enumerate}
\usepackage{amsmath}
\usepackage{float} 
\usepackage{amsfonts}
\usepackage{amssymb}
\usepackage{enumitem}
\usepackage{physics}
\usepackage{xcolor} 
\usepackage{graphicx}

\usepackage[left=1.2in,top=1in,right=1.2in,bottom=1in,headheight=0.8in,foot=0.5in]{geometry}
\usepackage[nodisplayskipstretch]{setspace} 
\usepackage{mdframed}

\usepackage[greek.polutoniko,english]{babel}

\hypersetup{
	colorlinks=true,         
	linkcolor=MidnightBlue,          
	citecolor=MidnightBlue,
	urlcolor=MidnightBlue            
 }
 \let\oldmarginpar\marginpar
\renewcommand\marginpar[1]{\oldmarginpar{\color{red}\raggedright\scriptsize #1}}
\let\oldmarginpar\marginpar
\renewcommand\marginpar[1]{\oldmarginpar{\color{red}\raggedright\scriptsize #1}}

\usepackage{natbib}
\setcitestyle{aysep={}} 

\title[\sc Theories Without Models]{\sc \large{Theories Without Models}: \\ \large{Uncontrolled Idealizations in Particle Physics}}
\date{Draft of \today.}

\usepackage{amsaddr}

\author{Antonis Antoniou}
\address{\vspace{-0.8pc}University of Bonn}
\email{\href{aanton@uni-bonn.de}{aanton@uni-bonn.de}}
\author{Karim P. Y. Th\'ebault}
\address{\vspace{-0.8pc}University of Bristol}
\email{\href{mailto:karim.thebault@bristol.ac.uk}{karim.thebault@bristol.ac.uk}}

\makeatletter
\let\uppercasenonmath\@gobble

\begin{document}
\setstretch{1.2}
\maketitle

\begin{abstract}
The perturbative treatment of realistic quantum field theories, such as quantum electrodynamics, requires the use of mathematical idealizations in the approximation series for scattering amplitudes. Such mathematical idealizations are necessary to derive empirically relevant models from the theory. Mathematical idealizations can be either controlled or uncontrolled, depending on whether current scientific knowledge can explain whether the effects of the idealization are negligible or not. 
Drawing upon negative mathematical results in asymptotic analysis (failure of Borel summability) and renormalization group theory (failure of asymptotic safety), we argue that the mathematical idealizations applied in perturbative quantum electrodynamics should be understood as uncontrolled. This, in turn, leads to the problematic conclusion that such theories do not have theoretical models in the natural understanding of this term. The existence of unquestionable empirically successful theories without theoretical models has significant implications both for our understanding of the theory-model relationship in physics and the concept of empirical adequacy.
\end{abstract}

\setcounter{tocdepth}{1}
\tableofcontents
\setstretch{1.4}

\newpage
\section{Theories Without Models: An Argument}\label{argument}

In this paper, we will articulate and explore a chain of reasoning encoded in the following argument pattern.  
\begin{enumerate}
\item [P1.] Perturbative quantum field theory is best understood as a \textit{framework for theories} where individual theories are picked out within the framework via (for example) their field content, symmetries, and actions.
\item [P2.] For a theory to have theoretical models is for a representation of the quantities predicted by the theory to be derivable either: i) as deductive consequences of the theory; or ii) as a result of an approximate derivation based upon \textit{controlled mathematical idealizations} within the theory.
\item [P3.] Contemporary particle physics features perturbative quantum field theories, such as quantum electrodynamics, in which we have good reasons to believe the derivation of predictions requires an \textit{ineliminable appeal to uncontrolled idealizations}.
\item [C1.] Contemporary particle physics includes theories, such as quantum electrodynamics, which we have good reasons to believe are theories without theoretical models. 
\end{enumerate}

P1 we take to be uncontroversial and straightforwardly justified by contemporary scientific practice in which quantum electrodynamics and quantum chromodynamics are treated as distinct theories within the framework of quantum field theory and not as models of quantum field theory. P2 is the natural definition of models of a theory in physical context and is consistent with that used in the philosophical literature \citep{vanFraassen1980,french:1999,sep-structure-scientific-theories}. We take it to be also  uncontroversial, although open to contestation and revision with good reason. The term `theoretical model' is used here as a term of art and its precise meaning will be clarified below. P3 is the controversial premise, and much of the paper will be devoted to defending it in the context of the well-known formal challenges to the rigorous formulation of perturbative quantum field theories. Given our arguments in favour of P3 are accepted, and one wishes to reject the conclusion, either on the grounds of being unintuitive or due to other negative implications, the obvious implication via contraposition is then that we should revise our definition of theoretical models to reject P2. 

We are sympathetic to such a response but note that such a revision will have non-trivial knock-on implications for the philosophy of science. In particular, changing our definition of a theoretical model to include theory-model links via uncontrollable idealizations will have implications for non-arbitrary connections between theories and models and the definition of empirical adequacy. We will consider these implications in the final Section \ref{implications}. There we will consider the idea of connecting empirically relevant models to a theory via reference to further theories or frameworks. That is, of modifying the definition of empirical adequacy to allow the models on which the adequacy is based to be linked to the theory via \textit{exogenous arguments}, which may include idealizations controlled via other theories or frameworks. This approach has the advantage of mirroring the attitude of many practicing scientists and has the interesting broader implication that perhaps philosophers should look beyond the theory as a unit of analysis in the context of evaluations of empirical adequacy. 

Before we proceed to situate our analysis in the various background literatures, let us make the aims of this paper completely clear by indicating what this paper is \textit{not} about. Here we are not arguing that particle physics does not have models \textit{per se}, nor that there is a fundamental or basic inconsistency or inadequacy in the framework of perturbative quantum field theory. Indeed, models of various non-theoretical sorts abound in particle physics and there exist various important mathematical results that indicate precisely in which circumstances the idealizations involved in perturbative quantum field theory can be understood to be controllable. 

This article serves two main goals. The first is to demonstrate that in the context of unquestionably empirically successful examples of individual perturbative quantum field theories applied in particle physics, such as quantum electrodynamics, the conditions for the controllability of the idealizations in question can be expected to fail. The second is to point out that this implies that these theories do not have models in the straightforward sense of theoretical model. We take this conclusion to be the result of the combination of  uncontroversial results in perturbative quantum field theory and straightforward definitions in the philosophy of science and, as such, to be all the more worthy of interest given the modesty of the individual argumentative steps involved. 

\section{Background}

\subsection{Mathematical Idealizations in Particle Physics}

Mathematical idealizations are formal simplifications of the structure of a mathematical model or system of equations that aim to facilitate the provision of explicit analytical or numerical solutions. The classic simple example of a mathematical idealization is the small angle approximation in the solution of the differential equation for pendulum motion and we will return to this example later. In modern physics, mathematical idealizations are often based upon the truncation of series expansions. For example, approximating the exponential function by the first few terms of the Taylor expansion, i.e. $e^x \approx 1+x+\frac{x^2}{2}$. Mathematical idealizations should be distinguished from other forms of idealizations, much discussed in contemporary philosophy, which typically focus on the processes of constructing model systems or representations by distorting or omitting features of their target systems.\footnote{In his seminal paper on  idealization \cite{McMullin1985} explicitly distinguishes between `mathematical idealization' and `construct idealization'. The discussions in \cite{Pincock2007} and \cite{Batterman2009} mainly concern the former type in which the focus is on the mathematical treatment of the models, while discussions in \cite{Weisberg2007}, \cite{Morrison2015} and \cite{Portides2021} mainly concern the latter. See also \cite{norton:2012} for related discussion.} 

A controlled idealization is an idealization for which all the necessary resources are available in current scientific knowledge to explain why the effect of the introduced assumptions in a model is, up to required degree, negligible.\footnote{Our deployment of this concept builds upon the idea of a  `controllable idealization' due to \citet[pp.44-5]{Sklar2000}. See also \cite{batterman:2005,Batterman:2014,wayne:2011,king:2016,knuuttila:2019}.} A controlled mathematical idealization is then a simplification of the formal structure of a mathematical model or system of equations in which all the necessary resources are available in current scientific knowledge to explain why the effects of the relevant formal simplifying assumption are, up to a required degree, negligible. Such resources may come either from within the theoretical tools of a given background physical theory in which the idealizations occur, or, when this is not possible, by appeal to different physical theories in which case one has what we shall call an exogenous justification. 

Examples of mathematical idealizations in physics that, when all goes well, are controlled include linearity assumptions, infinite differentiability assumptions, and adiabaticity assumptions (i.e. one function is assumed to change slowly with respect to another). Such idealizations are controlled when scientists are able to provide reasons  why the effects of the relevant formal simplifying assumption are, up to a required degree, negligible. Typically, in controlling a mathematical idealization a scientist is able to provide estimates of the deviation between the theoretical predictions of the idealized model and the exact solutions of the original mathematically intractable theoretical model. 

In what follows, we will distinguish between the situation where we do not at present have the resources to control an idealization but we have good reason to expect that this is only a practical limitation and the situation where we have good reason to believe that we will never be in the situation to control the idealization in question. The first we will call \textit{uncontrolled} idealizations and the second \textit{uncontrollable} idealizations.

The standard model of particle physics is a marvel of empirical science and has been variously tested to astounding degrees of accuracy. The framework within which the standard model is presented is perturbative Quantum Field Theory (QFT).\footnote{There have been various attempts to cast the standard model in terms of non-perturbative quantum field theory. However, to date, such attempts have been met with limited success. Here we focus exclusively on perturbative approaches with the framework of Lagrangian Quantum Field Theories and neglect analysis of models in Algebraic Quantum Field Theory (AQFT) approaches. For the classical philosophical discussion of the relative merits of perturbative and algebraic approaches see \cite{Wallace2006,fraser:2011,fraser:2009,wallace:2011}. An issue of particular interest, worthy of consideration in future work, is the implications for our analysis of work on perturbative approaches to AQFT \citep{rejzner:2016}.} Following \cite{Wallace2006, Wallace2018} and \cite{Koberinski2021} our analysis is based on an understanding of perturbative QFT as a general theoretical framework of principles and general mathematical constraints from which certain well-defined Lagrangian quantum field theories describing particular interactions are constructed, such as quantum electrodynamics (QED) and quantum chromodynamics (QCD).\footnote{Here we deviate slightly from the terminology of \cite{Koberinski2021}, since he prefers to refer to the Lagrangian QFTs as dynamical models, although he admits that he uses the term `dynamical model' in the way most would use the term `theory'.} These Lagrangian QFTs necessarily come with an implicit cut-off limit implying that they are effective field theories describing phenomena within a particular range of energy scales. The connection of these theories with the empirical data is achieved via the construction of empirically relevant models of specific interactions within the theory's domain (e.g., a meson-meson scattering for QED), usually by perturbation methods. In what follows, quantum electrodynamics is thus referred to as the `background theory' from which empirically relevant models of certain interactions are derived with the use of approximation techniques.

The perturbative treatment of the class of realistic models in quantum field theories requires an approximation of the scattering amplitudes in terms of a divergent power series of the coupling parameter. The divergence occurs (i) from the fact that the total sum of the series is not known to converge or known not to converge and/or (ii) from the fact that each individual term of the series diverges at very high and very low energies leading to the so-called \textit{ultraviolet} and \textit{infrared divergences} respectively. As will be shown, the necessary mathematical treatment of these challenges in the case of important examples of QFTs, such as quantum electrodynamics, requires the use of \textit{uncontrolled idealizations} which break the formal link needed to identify the background theory with its theoretical models. In particular, in cases where the corresponding functions are unknown like in perturbative QED, the justification for the truncation of a series expansion after the first few terms can be established if the series is \textit{strongly asymptotic}, by showing that it is Borel summable, cf. \cite{Miller:2023}. Correspondingly, the justification for the renormalizability of a quantum field theory via renormalization group methods can be achieved by showing that a theory is \textit{asymptotically safe} by calculating the corresponding beta-function, cf. \cite{Huggett2002}. However, these requirements have not been demonstrated to hold in important examples of QFTs, and there is evidence that they do not in fact hold in QED at least. Thus, the introduced mathematical idealizations are uncontrolled, and plausibly they should be understood as uncontrollable.

It is important to note, in this context that the formal challenges regarding the establishing of the renormalizability of a quantum field theory are \textit{not}  ameliorated by appeal to the Effective Field Theory (EFT) framework and the formal control such a framework offers over perturbatively non-renormalizable terms. Indeed, the `Wilsonian' renormalization group approach, which we will consider in detail later, is one of the principal formal foundations of the EFT paradigm. Moreover, the other key formal foundation, and the basis of the crucial property of `decoupling' between effective and high-energy theories, is the Appelquist-Carazzone theorem \citep{appelquist:1975} and this assumes an underlying renormalizable theory (with different mass scales) \citep{hartmann:2001}. Furthermore, following the recent arguments of \cite{Franklin2020}, \textit{it is (effective) renormalizability that accounts for the effectiveness of effective field theories}. The EFT framework is an illustration of rather than an alternative to the need for justification for renormalizability in QFT.

Given the inability to control the idealizations within the theory, a possible route towards the justification of the mathematical idealizations of QED is to look to physics beyond the standard perturbative analysis and particle content (e.g. introduce `instanton' effects to remedy the failure of Borel summability) or beyond the standard model entirely (e.g. consider implications from asymptotically safe quantum gravity theories to deal with the failure of asymptotic safety). While this is a fruitful way of controlling idealizations in physics, when such exogenous justifications appeal to different physical theories, they cannot establish that the resulting models are, strictly speaking, models of the theory of QED. As will be shown in the next subsection, under the natural sense of `models of a theory' relevant to physics, a theoretical model is a model that can be deductively derived from the background theory modulo some controllable idealizations \textit{within the theory} -- i.e. using resources from the tools of the theory plus mathematics and logic. QED, as it currently stands, is unable to provide its own models due to the inability of controlling the idealizations involved. Most significantly, even if the establishment of a UV completion within a quantum gravity scenario ultimately turns out to be successful, the new resulting models will ultimately not be models of QED, but rather models of the conjunction of QED and the new theory of quantum gravity, given that all the introduced idealizations are controllable.\footnote{This feature points to a crucial disanalogy between the notion of a `model of a theory' as understood in the context of physical theories and as applied in model theory. Whereas, in model theory if $M$ is a model of theory $T+T'$ it is also a model of theory $T$, such situations do not in general obtain in physical theory. Most explicitly, we could consider $T$ to be special relativity and $T'$ to be general relativity. There are then clearly models of  $T+T'$ which are not models of $T$. Thanks to an anonymous reviewer for pointing this issue out to us.} This is orthogonal to our conclusions, however. As it stands, and when understood as a theory of the interactions of charged particles with the electromagnetic field, QED does not provide theoretical models in the natural sense of `models of a theory' relevant to physics. We shall return to this issue in Section \ref{cutting}.

The general understanding of perturbative quantum field theory that we adopt is close in spirit to earlier discussions on the role of theories, models, approximations and renormalization techniques in quantum field theories. We conclude this subsection by giving a brief overview of this literature. 

To our knowledge, the first philosophical discussion of the specific problem of justifying the deductive relations between theories and models in QFT can be found in the wide-ranging and hugely insightful article of  \cite{hartmann:2001}. Based upon a detailed discussion of QCD and EFTs the author notes that `there might not be a `controlled' deductive relation between the model and an underlying theory, and if a derivation of the model from the theory is actually carried through, further assumptions have to be made to obtain the model, and these assumptions (which might turn out to be more dubious than the assumptions made by the original model) again require a justification, and so on, \textit{ad infinitum}' (p. 293). We take this analysis to be very much in line with our own appraisal of the theory-model relation for the case of QED. 

In a more recent analysis, related to but distinct from our own, James Fraser argues that perturbative methods in quantum field theory do not produce models, rather they only produce approximations of certain physical quantities such as cross-sections and decay widths \citep{Fraser2020} . In more recent work, Fraser's view on the role of approximations in quantum field theory has been further elaborated together with co-authors in \cite*{deOlano2023} by showing, via certain historical examples, that approximations are instrumental in determining the empirical content of perturbative quantum field theories and should be recognised as a distinct and equally important category of theoretical output compared to idealized systems. \cite*{deOlano2023} further stress that the integral role of approximations in assigning empirical and physical content to a model cannot be easily accounted for by extant philosophical approaches to scientific modelling. 

\cite{Koberinski2021} also provides a historical study of the mathematical developments of Yang-Mills theories to highlight the fact that quantum field theories are rarely axiomatic systems with a neat set of deductive consequences, but are rather modular and rely heavily on key conceptual and mathematical tools that can be treated independently. Similarly, \cite{Miller:2023} argues that the state-space semantics developed by \cite{Beth1960} and \cite{vanFraassen1970} fails to capture perturbatively evaluated observables in cases of divergent series, and proposes a slight modification that allows the accommodation of these results. These results complement and support our own approach. As such, we take ourselves to be operating in the mainstream of the philosophy of quantum field theory, despite the counter-intuitiveness of our conclusion.

\subsection{Models and Theories}

Under what might be called the `canonical view' on theoretical models, to be a model of a theory is to stand in the relation to the theory that is analogous to that between a logical model and a formal theory expressed as a set of axioms or formal sentences.  Such a view of theoretical models is common to most major approaches to the structure of scientific theories. That is, whether or not one holds the semantic view of theories under which theories are identified with the theoretical models, it is still standard to consider the models of a theory as an important aspect of a theory's structure and to define such theoretical models via the canonical view.\footnote{See, for example, \cite{savage:1990,french:1999,lutz:2012,sep-structure-scientific-theories}.} The natural sense of `models of a theory' that is relevant to physics draws on an analogy to the use of model in model theory but is a clear and independent conceptualisation.\footnote{We note that some approaches to the semantic view invest the analogy with deeper significance.} The most vividly and logically rigours example is the models of general relativity are unambiguously specified by pairings of stress-energy tensors and Lorentzian four metrics that solve the Einstein Field Equations. A highly influential articulation of the canonical view is due to \cite{vanFraassen1980}, who indicates that theoretical models should be understood as deductive products of the background theory in the sense that they are derived from a set of fundamental equations: `[t]he sense in which a theory offers or presents us with a family of models -- the theoretical models -- is just the sense in which a set of equations presents us with the set of its own solutions [...] When the equations formulate a scientific theory, their solutions are the models of that theory' (\textit{ibid.} p.311). 

Significantly, just like mathematical equations, a theory may have models that we have not yet discovered or fully articulated, but nonetheless exist. Furthermore, again just like mathematical equations, a theory may be associated with models which have not been demonstrated to be deductive consequences of the theory but rather to be under a set of simplifying assumptions. Using the terminology of the last section, one can understand as theoretical models those models whose deductive relation to the theory is based upon a controlled mathematical idealization. We will articulate this idea further in the following section. For the time being the important point is that in the context of  examples of QFTs, such as quantum electrodynamics, where the theory-model association is based upon mathematical idealizations that are uncontrolled, there is no longer a clear sense in which the models in question are models \textit{of the theory}. 

We thus end up with a deeply unintuitive situation where we have theories without models. On close inspection the situation turns out to be not just unintuitive but also concerning since theoretical models can be expected to play a number of important roles within scientific theories. A feature of particular significance is that on most accounts it is the theoretical models that allow theories to make contact with empirical evidence. This feature is not only pivotal to van Fraassen's constructive empiricist view of science but can be expected to be central to any account of how scientific theories are applied in practice. In particular, the \textit{empirical adequacy} of a theory is standardly understood to be assessed by the predictive success of the theoretical models it provides. Thus, we arrive at the significant worry that the existence of theories without models not only undermines the pristine picture of theories and theoretical models, but also indicates that the empirical adequacy of theories such as quantum electrodynamics cannot be cogently expressed. We will return to this worry in Section \ref{implications}.

The autonomy of models from theories as a criticism to the semantic view of theories has of course been discussed in the past in other contexts.\footnote{For instance, \cite{Portides2005} criticises the semantic view by showing how the liquid drop model of the nucleus was introduced in the 1930s to explain -- amongst other things -- the Weizsäcker semi-empirical formula of the binding energy of the nucleus. Portides convinsingly argued that certain crucial assumptions of the model lack the necessary theoretical justifications which would allow one to consider the liquid drop model as a theoretical model (\textit{ibid.}, p.1294). In a similar spirit, \cite{Cartwright1995} and  \cite{Morrison1999} argued for the autonomy of models from theories using the examples of the Londons's model of superconductivity and  Prandtl's model of ideal fluids respectively.} However, such arguments typically appeal to examples of \textit{phenomenological models} in physics. The main characteristic of phenomenological models is that their construction is primarily guided by empirical observation and experimental data, maintaining minimal dependence on the background theoretical framework. As a result, such models often violate basic theoretical principles and fundamental laws of the background physical theory. Phenomenological models are thus a type of non-theoretical models in that they are independent and autonomous from the theory. Our worry is rather different. That is, our claim is that there is a very different type of non-theoretical models in physics -- for which the models of perturbative quantum field theory are a prime example -- in which the relationship between theories and models breaks down not because these models are primarily guided by phenomenological considerations; but rather, because the mathematical treatment of these models requires the introduction of uncontrollable idealizations due to the lack of a rigorous mathematical or physical justification. 

The class of problematic models that we identify is similar to what \cite{Redhead1980} calls \textit{floating models}. For Redhead, these are models that are disconnected from a background theory due to a \textit{computational gap} arising from the scientist's inability to justify the validity of the mathematical approximations used for their construction, but at the same time, unlike phenomenological models, fail to provide successful empirical predictions. They are thus, in a sense, `floating' in that they are detached both from theory and the experimental data. Their value is to be found in the fact that they often serve as preliminary steps for probing various essential features of the theory and exploring the possible ways of refining it in order to yield further empirically successful models. However, unlike Redhead's floating models, the models of perturbative quantum field theory exhibit remarkable agreement with experimental data despite the fact that they  are not connected to the background theory via a secure approximate derivation based upon a controlled mathematical idealization. This is, we contend, a deeply challenging situation that has not been adequately addressed in the vast literature on theories and models in science. 

\section{Theoretical Models and Empirical Adequacy}\label{Theoretical}

In this section we will specialize our analysis of theoretical models to the particular case of van Fraassen's constructive empiricism and consider the relationship to empirical adequacy and mathematical idealizations in that context. As noted above, we take our analysis to generalise to other approaches, and thus the reader should see our work here as illustrating the general features of the canonical view on theoretical models and the relationship to empirical adequacy.  

According to van Fraassen, in articulating a theory one begins with a set of fundamental principles and laws and, without taking into consideration any kind of empirical data, constructs a set of models which can be deductively derived from these principles. These models are then compared to the `appearances', i.e., the structures which can be described in experimental and measurement reports. A physical theory is empirically adequate `if it has some model such that all appearances are isomorphic to empirical substructures of that model' \citep{vanFraassen1980}. van Fraassen's emphasis on the importance of theoretical models of a theory stems from his strong views in favour of the semantic view of the structure of scientific theories where the focus is on models, rather on the linguistic formulation of theories.\footnote{cf. \citet[p.217]{vanFraassen1989}:`[The semantic approach's] difference from other approaches is largely one of attitude, orientation, and tactics rather than in doctrines or theses. The conviction involved is that concepts relating to models will be the more fruitful in the philosophical analysis of science'.} For \cite{vanFraassen1980,vanFraassen1989}, the presentation of a scientific theory consists of a description of a class of state-space types, where state spaces are understood as a generic collection of mathematical objects (such as vectors, functions, and numbers) denoting the physical state of a system. Theoretical models are understood in this context as mathematical structures that are deductively derivable from the basic principles of the theory, denoting sequences of states that form a trajectory in the state space over time. The connection of these models to the physical world is described by \citet[pp. 168-170]{vanFraassen2008} in terms of an isomorphic relationship between the theoretical models and the surface models of an experiment, i.e. the idealized models of the raw data obtained from experiments.\footnote{See \cite{lutz:2014,lutz:2021} for interesting work on generalizing van Fraassen's notion of empirical adequacy. } Together with the observable phenomena, theoretical models occupy the central stage of constructive empiricism, since these are `the two poles of scientific understanding, for the empiricist [..] The former are the target of scientific representation and the latter its vehicle' \citep[p. 238]{vanFraassen2008}.\footnote{Things are of course not so straightforward when it comes to quantum and relativistic theories. For more details on van Fraassen's treatment of these cases see \citet[Ch.6.3]{vanFraassen1980} and \citet[Ch.5]{vanfraassen:1991}.}
	
A characteristic example of a theoretical model in physics is the ideal pendulum in Newtonian mechanics. The dynamic behaviour of the system described in this model is completely determined by the equations of the background theory and is in full accordance with the theory's fundamental principles. However, this purely theoretical model does not allow the (straightforward) comparison of the model with the corresponding surface models of data derived from experiments. Hence, the additional construction of \textit{empirically relevant models} -- based on these theoretical models -- is often necessary, to compare the quantities predicted by the models with the outcomes of experimental measurements. Theoretical models thus provide the mathematical basis on which further empirically relevant models are built with respect to a corresponding target system. The aim is to obtain theoretical predictions for experimental observables via these new models, in order to test the empirical adequacy of the models and consequently of the background theory.

As van Fraassen notes (\textit{ibid.} p.311), the process of arriving at a working model from a background theory can take several forms, often involving the use of mathematical idealizations. These idealizations typically occur in cases where the theory provides theoretical models for which no analytic solutions can be found and thus, further modifications are required in order to make the models mathematically tractable. For instance, the (theoretical) model of the ideal pendulum in classical mechanics provides is given by the equation of motion
\begin{equation}\label{eq:pend}
	\dv[2]{\theta}{t}+\frac{g}{l}\sin\theta=0
\end{equation}
where $\theta$ is the angle between the string and the vertical, $g$ is the local gravitational acceleration and $l$ is the length of the string. Equation (\ref{eq:pend}) specifies, in principle, the aforementioned stochastic response function according to which a measurement of $\theta$ will yield a specific value $\theta_1$ given a state $s_1$, and is a second order non-linear differential equation for which exact analytic solution requires advanced methods of elliptic functions. However, the introduction of a mathematical idealization allows for a mathematically simple empirically relevant model which will eventually enable the comparison to the relevant surface model.

This is achieved by introducing the assumption that the pendulum only swings in small angles for which we know that
\begin{equation}\label{sin}
	\sin\theta\approx\theta
\end{equation}
The small angle approximation yields a new linear equation of motion	\begin{equation}\label{eq:pendlin}
	\dv[2]{\theta}{t}+\frac{g}{l}\theta=0
\end{equation}
which has a simple exact solution that can be derived via basic approaches to solving differential equations. Insofar as there are good mathematical reasons to justify that (\ref{sin}) holds for small angles, the introduced idealization is controllable and the linearized version of the ideal pendulum is still a theoretical model, in that it is derivable from the theory modulo some controlled modifications.

In what follows, we draw on the modelling techniques of perturbative quantum field theory to argue that in addition to phenomenological models whose construction aims in accommodating empirical results, there is another important class of non-theoretical models in physics whose construction comes from the use of non-controllable idealizations aiming to make them mathematically tractable and empirically relevant. A careful analysis of the modelling methodology of quantum electrodynamics within perturbative quantum field theory indicates that the empirically relevant cut-off models produced by regularization and renormalization techniques are detached from their corresponding theoretical models and the background theory, due to the fact that the required justification for the truncation of the series and the employment of the renormalization group method has not been established in practice. This fact breaks down the derivational relationship between the empirically relevant models and their background theory, leading to problems both for our understanding of the theory-model relationship and the empirical adequacy of the theory according to van Fraassen's view and more widely. 

\section{Models in Perturbative QFT}\label{QFTmodels}
	
Perturbative Quantum Field Theory is the standard theoretical framework for constructing models for scattering processes of subatomic particles in particle physics. Realistic models of interactions are  constructed by performing scattering theory calculation using specific quantum field theory Lagrangians or Hamiltonians, such as those for QED and QCD. These calculations allow the construction of empirically relevant models that are capable of providing theoretical predictions for physical quantities that can then be compared directly to experimental observables, such as cross sections and decay widths. A first difficulty in this process arises from the fact that the construction of four-dimensional interacting models via scattering theory leads to intractable Hamiltonians since the introduction of interacting terms typically leads to non-linear equations of motion for which no exact solutions can be found analytically.
	
In order to make the interacting models mathematically tractable, physicists often appeal to perturbation theory, in which the total Hamiltonian $H$ of the system in question is assumed to be well-defined and equal to the sum of a free and an interacting part:
\begin{equation}\label{Hamiltonian}
	H=H_0+gV
\end{equation}
where $H_0$ is the free Hamiltonian whose eigenstates can be calculated analytically and $gV$ represents the Hamiltonian of the interacting model with an interaction potential $V$ parameterised by a coupling parameter $g$. Given that the coupling strength of the theory is weak -- as is the case in QED and in high-energy QCD -- this mathematical idealization is controllable, since it can be justified by safely assuming that the solutions of the total Hamiltonian are close enough to the well-known solutions given by the free Hamiltonian $H_0$.
	
One of the most important quantities to be obtained by applying perturbation theory is the S-matrix, the operator that maps the initial state of a physical system undergoing a scattering process to the final state. 
The importance of the S-matrix for particle phenomenology experiments boils down to the close relationship of its elements -- also referred to as scattering amplitudes $\mathcal{M} $ -- to the scattering cross sections that are (indirectly) measured in these experiments. The S-matrix thus contains some of the theoretical predictions of the empirically relevant model of the theory upon which the empirical adequacy of the model can be tested. 
	
When perturbative methods are applied to realistic quantum field theories like QED, the final result is an expression of the scattering amplitudes as a power series of the (bare) coupling parameter $g_0$ which has the following generic form
\begin{equation}\label{series}
	\mathcal{M}=\sum_{n}^{\infty}{g_0}^n\int_{-\infty}^{\infty}dkA_n
\end{equation}
where $A_n$ stands for a multiple integral over momentum $k$ at each $n^{th}$ order term of the series. For those interactions where the coupling constant is relatively small, the first factor ${g_0}^n$ vanishes as the series proceeds to higher orders, and thus the contribution of higher order terms decreases significantly. All that is needed to calculate the coefficients of the series for each order (and consequently the S-matrix elements) is, therefore, to evaluate a set of multiple integrals over momentum space.
	
This is where two significant problems arise. The first problem concerns the fact that the total sum of the series is sometimes either not known to converge or known not to converge, and thus a legitimate question arises as to whether the series is indeed a good approximation to the quantity in question. Moreover, the number and the complexity of each of these integrals increases rapidly for higher orders and thus, in practice, only the sum of the first few terms is evaluated and compared to experimental results, often with great success. The second complication stems from the fact that each individual term of the series often appears to diverge either at very high energies (as $k\rightarrow \infty$)  or at very low energies (as $k\rightarrow 0$) leading to the so-called \textit{ultraviolet} and \textit{infrared divergences} respectively. This is the notorious `\textit{problem of infinities}' in quantum field theory and, as we shall see, the solution to this anomaly comes from various regularization and renormalization techniques. 
	
The crucial question is whether the mathematical methods employed at this stage can be rigorously justified, making the introduced idealizations controllable, as in the simple case of the ideal pendulum. If so, then one is warranted to maintain that the final products, i.e. the empirically relevant models that are eventually put to test by comparison with experimental results, are theoretical models by virtue of being derivable from the background theory via controllable idealizations. Here is where we encounter a problem however, since the mathematical treatment of these two complications indicates that the introduced idealizations are \textit{uncontrolled and possibly uncontrollable} due to the lack of a rigorous mathematical proof for (i) the Borel summability of the series to establish \textit{strong asymptoticity} and (ii) the calculation of the corresponding beta-functions in renormalization group methods to establish \textit{asymptotic safety}. As a result, the strong relationship between the final models and the background theory required by the semantic view breaks down. Let us consider each of these problems in more detail.
	
\section{Breaking the Theory-Model Link}\label{cutting}
	
\subsection{Divergent Series}

The first complication concerns the large order behaviour of the perturbative series as a whole, independently of the offending integrals in the individual terms. In addition to the fact that the integrals in each term of the series diverge, there is often no evidence that the total sum of the series converges, or, to make things worse, there is strong evidence that the sum of the series at all orders diverges. Given the infinite number of terms and the increasing intractability of the integrals as one proceeds in higher-order terms, what happens in practice is that only the first few terms of the series are usually calculated and compared to experimental results, often by using different techniques for calculating each term of the series. Nevertheless, in several cases the sum of the first few series turns out to be in very close agreement with the experimental results, which leads us to the question of how the empirical success of the truncated series should be explained.\footnote{The theoretical prediction for the anomalous magnetic moment of the electron is probably the most famous example of this practice. The first term of the series was first calculated by \cite{Schwinger:1948} and found to be in close agreement with the then available experimental results. The current state of the art only allows the calculation of the first five terms which requires -- amongst other things -- an evaluation of 12672 Feynman diagrams in the tenth-order perturbation theory \citep{Kinoshita2014}.}
	
\cite{Miller:2023} discusses at length the case of divergent series in perturbative quantum field theories and notes that a possible justification for the truncation of the series after the first few terms and the explanation of the empirical success of the approximating series can be given in terms of \textit{strong asymptoticity}.\footnote{As Miller notes, this explanation was also given by \cite{Dyson1952} himself in his arguments for the divergence of the perturbative series.} Unlike convergent series, when a divergent asymptotic series is used for approximating a function, the sum of the first few terms is very close to the exact value of the function, and as one includes more and more terms to the sum, the value of the series increasingly diverges until it becomes infinitely different. Hence, if the perturbative series for the matrix elements of a meson-meson scattering process is indeed asymptotic to the unknown exact solution, it should be no surprise that the sum of the first few terms is often found to be in agreement with experimental results.
	
The problem, however, is that while a given function can only have one asymptotic expansion, the converse is not true. That is, an asymptotic expansion can correspond to multiple functions, and thus it is possible that two or more different functions have the same asymptotic expansion. In fact, it is possible for an infinite number of functions with an entirely different set of solutions to share the same asymptotic function. Hence, even if a perturbative expansion in QED is indeed asymptotic to an exact solution, it does not uniquely specify what that function is. To uniquely determine a function via an asymptotic series one shows that the series satisfies a \textit{strong asymptotic} condition, which, roughly speaking, requires the differences between the exact value of the function and its series representation to be appropriately small for every order of perturbation theory. 

The standard way to establish the condition of strong asymptoticity that shows that a perturbative expansion uniquely satisfies an exact solution is by showing that a series is \textit{Borel summable}. Borel summation is a mathematical method that is particularly useful for summing divergent asymptotic series. Suppose a formal power series $\sum_{k=0}^{\infty}\alpha_k\lambda^k$ of a function $f(\lambda)$ and define the \textit{Borel transform} $B$ to be its equivalent exponential series $B(\lambda t)\equiv\sum_{k=0}^{\infty}\frac{dk}{k!}t^k$. If the Borel transform can be used to produce a unique reconstruction of $f(\lambda)$ then the series is Borel summable. The important point for our purposes is as follows: If one knows the asymptotic expansion of a function, the function can be uniquely reconstructed by Borel summation, thus showing that the series corresponds to a unique function determined by the background theory. Unfortunately, however, Borel summability has \textit{not only not been demonstrated} but is in fact \textit{not expected to hold} in  phenomenologically interesting models of quantum field theory. Indeed, \cite{Duncan2012} notes that `the property of Borel summability is an extremely fragile one, and one which we can \textit{hardly ever} expect to be present in interesting relativistic field theories' (p.403, emphasis added). For perturbative QED in particular, Borel summability is not expected to hold in general.\footnote{Here we are following the discussion of \cite{Miller:2023}. For further discussion in the physics literature see \cite{thooft:1979} \cite{khuri:1981,fischer:1997}.} 

It should be noted however, that Borel summation is not the only available reconstruction technique for connecting perturbative expansions with exact solutions.\footnote{For a more rigorous discussion on Borel summation and possible alternatives in the context of perturbative QFT see \cite[pp. 400-6]{Duncan2012}.}  In fact, an alternative group of approaches under the name of `optimized perturbation theory' retains some hope for potentially associating the asymptotic series with a unique function. An example of these methods is `linear $\delta$ expansion' where the basic idea is to construct a series of approximates to the path integral which only require perturbative calculations, but nevertheless can be shown to converge rigorously to the exact answer. Roughly speaking, this technique can, in principle, establish that a sequence of approximations obtained by carrying out the $\delta$ expansion converges to an exact answer even if the original asymptotic expansion is not Borel summable. The problem however is that in the case of quantum field theories the radius of convergence is essentially unknown, and higher loop calculations are simply intractable so one is left to hope for a rapid convergence at low orders, usually less than five.

The upshot is that no available mathematical technique has thus far succeeded in associating the series with a unique solution. Thus the explanation for the success of the series after the truncation of the first terms based on strong asymptoticity is only a conjecture. The relevant mathematical idealization is therefore uncontrolled and the derivational relationship between the empirically relevant models in perturbative QFT and the background theory is, as it stands, obstructed. We will return to this issue and consider the idea of \textit{exogenous} control of the idealization via non-perturbative `instanton' effects in Section \ref{implications}.

\subsection{Cutting Off Our Ignorance}
	
The second problem that needs to be addressed in the construction of empirically relevant models in perturbative QFT is the presence of ultraviolet and infrared divergences in the individual terms of the power series in (\ref{series}) which makes the calculation of the coefficients impossible. The elimination of these infinities is achieved by various methods of \textit{regularization} in which a divergent integral $A_n$ is redefined as a function of a new parameter $\xi$ -- the regulator -- which must satisfy the two following \textit{pragmatic constraints}: (i) finite values of the regulator must render the integral $A_n(\xi)$ finite and (ii) if the regulator is removed by taking its limit to infinity the retrieved result is the original divergent integral $A_n$.
	
A standard regularization method is the Pauli-Villars regularization in which a cut-off limit $\Lambda$ is introduced for the ultraviolet and infrared domains of high and low energies respectively, beyond which the value of the integral is taken to be zero.\footnote{Another popular method of regularization is dimensional regularization where, roughly speaking, the calculations are initially carried out in a d-dimensional spacetime, and the cut-off dependencies appear as we take the well-defined limit $d \rightarrow 4$. The Pauli-Villars method discussed here has the advantage of being physically more transparent as opposed to dimensional regularization which is considerably more abstract in nature. Some less popular methods of regularization are the lattice regularization, the zeta function regularization and the causal regularization.} In the case of ultraviolet divergences, this is equivalent to the introduction of a mathematical idealization according to which
\begin{equation}
	\sum_{n}\int_{\Lambda}^{\infty}A_n\equiv 0
\end{equation}
The introduction of the cut-off limit $\Lambda$ should be understood here as a physical construct introduced \textit{by fiat}, representing  \textit{our ignorance} about the validity of our methods beyond this limit. It is not a mathematically or physically justified constraint similar to $(\ref{sin})$ for instance. Rather, the only available justification for the introduction of the cut-off limit is \textit{instrumental}, in that it aims to make the calculations of the series' coefficients possible. In other words, we simply ignore contributions to the integrands for values of momentum greater than $\Lambda$ in order to remove the infinities and make the integrals physically relevant. A similar technique is used for the removal of the infrared divergences in the low momentum domain that typically occur in theories with massless particles, such as the photons of QED. In these cases, an infrared cut-off limit is introduced for a small but non-zero value of momentum below which the contribution of the integrands to the sum is neglected.\footnote{For a philosophical discussion on the treatment of infrared divergences see \cite{Miller2021}.} These methods eventually remove the undesirable infinities from the first few terms and thus the series provides an estimate of the quantity in question -- e.g. the S-Matrix element -- up to the order for which the infinities have been removed.\footnote{For textbook style expositions of regularisation and renormalization methods in quantum field theory see \cite{Peskin1995}, \cite{Zee2010} and \cite{Duncan2012}. For expositions of these methods aiming particularly at philosophers see \cite{Butterfield2015}, \cite{Wallace2018} and \cite{Williams2018}. For a philosophically rich historical discussion see \cite{fraser:2021}.}
	
An immediate implication of regularization is that the approximate value of the scattering amplitudes $\mathcal{M}$ now depends on the arbitrarily chosen value of $\Lambda$. However, given that there are no particular reasons for choosing a specific value for $\Lambda$ over a different one, say $\Lambda^2$, this should be worrying; different values of the arbitrarily chosen regulator will simply give different results for the scattering amplitudes, which is of course unacceptable. The scattering amplitudes provided by these models express the empirical content of our theory and thus they should not be a function of the theorists' arbitrarily chosen value for $\Lambda$.\footnote{It should be noted here that the choice of $\Lambda$ is not always completely arbitrary. For instance, when we want to abstract away from the interactions of a heavier, higher-energy particle, $\Lambda$ is chosen to be well below the mass of that particle and above the mass of the particles we are interested in studying. However, given that the range between the two masses is significantly large, the choice of the exact value for $\Lambda$ is still, more or less arbitrary. \cite{Butterfield2015} also point out that in some cases the background theory \textit{hints} towards a range of values for the cut-off limit that are plausible to take (p.448). An example of such a suggestion for the cut-off limit comes from the Lamb shift in QED, which suggests the electron's Compton wavelength as a natural lower limit for distance $d$. Nonetheless, even in these cases no rigorous mathematical or physical justification can be given and thus the final choice of $\Lambda$ remains arbitrary.}
	
The solution to this problem came from \textit{renormalization}, cf. \cite{fraser:2021}. In this method the unwanted $\Lambda$-dependence of the integrals, and consequently of the scattering amplitudes, is eliminated by extracting the value of the coupling parameter $g$ from scattering experiments at a particular energy scale $\mu$. Once the value of the coupling parameter is obtained, the original `bare coupling' $g_0$ is replaced by the new renormalized value $g_R(\mu)$ (sometimes called the `physical coupling parameter') which is now a function of the energy scale $\mu$ at which the experiment was performed.
	
As a concrete example, consider the perturbation series for the scattering amplitude of a meson-meson scattering process up to the second order correction. Once the ultraviolet cut-off is introduced, the amplitude $\mathcal{M}$ becomes a finite `cut-off dependent' quantity of the form
	
\begin{equation}\label{mesonmesonpseries}
	\mathcal{M}= -ig_0 + iC{g_0}^2[\log(\frac{\Lambda^2}{s})+\log(\frac{\Lambda^2}{t})+\log(\frac{\Lambda^2}{u})]+\order{{g_0}^3}
\end{equation}
where $C$ is a numerical constant, and the kinetic variables $s, t$ and $u$ are functions of the square of the energy at which the particles are scattered, and are related to rather mundane quantities such as the centre of mass energy and the  scattering angle. Once the value of the coupling parameter is experimentally obtained, the bare coupling parameter $g_0$ in (\ref{mesonmesonpseries}) is replaced by the renormalized coupling parameter $g_R$. By carrying out some simple algebraic calculations one is left with a new `cut-off independent' expression
	
\begin{equation}\label{mesonrenormalized}
	\mathcal{M}= -ig_R + iCg_R^2[\log(\frac{s_0}{s})+\log(\frac{t_0}{t})+\log(\frac{u_0}{u})]+\order{g^3}
\end{equation}
where the values of $s_0, t_0$ and $u_0$ are defined by the particular energy at which the coupling parameter $g_R$ was measured. At this stage, the dependence on the cut-off limit $\Lambda$ is dropped out, however, as one may notice the scattering amplitude now depends on the ratio between the energy scale $\mu$ at which the value of $g_R(\mu)$ is measured (captured by $s_0, t_0$ and $u_0$) and the energy scale of the future scattering experiments by which the empirically relevant model will be tested (captured by $s, t$ and $u$).
	
What is so special about the chosen energy scale $\mu$ however? And what if the renormalized parameter $g_R(\mu)$ was measured at a different scale $\mu'$? One might think here that this is not much of an improvement since the dependence on the arbitrarily chosen value of $\Lambda$ has simply been shifted to a dependence on the energy scales $\mu$ at which we are able to conduct experiments and extract the value of the renormalized coupling parameter. The standard way to deal with this situation in the context of condensed matter physics is to think of Quantum Field Theory as a theoretical framework for providing Effective Field Theories that only describe physical phenomena at a particular range of energy scales (up to $\mu$) and not as a fundamental theory of physics. This idea leads us naturally to the framework of renormalization group (RG) theory \citep{Wilson1971,fisher:1998}.
	
The renormalization group method provides the mathematical apparatus for a systematic investigation of the changes in the couplings of a theory with respect to the changes in the energy scales. Leaving the technical details aside, the main idea of this modern approach to renormalization is captured by the renormalization group flow equation
\begin{equation}\label{rgeq}
	\mu\derivative{}{\mu}g_i(\mu)=\beta_i(g_i)
\end{equation}
which determines the so-called \textit{beta function} $\beta_i(g_i)$ as the differential change of the running coupling parameters $g_i(\mu)$ of a QFT with respect to a small change in energy scale. If the theory happens to have several coupling constants $g_i, i=1, ... , N $, then the beta function is a function of all of the couplings in the theory $\beta_i(g_1,...g_N)$. One can think of the renormalization group equation as defining an N-dimensional space of theories in which $(g_1,...,g_N)$ are the coordinates of a particle that \textit{flows} in the space as the energy scale $\mu$ increases. In the high-energy regime where we have no experimental access to measure the values for the running coupling parameters $g_i(\mu)$, there are four possible behaviours of the theory's couplings depending on the beta function:
	\begin{enumerate}[label=(\Roman*)]
		\item \textbf{Safety}. At a finite flow parameter value the renormalization group flow hits a point (also known as an attractor) at which the beta function becomes zero for all running couplings, $g_i(\mu)$. This is the so-called fixed point $g*$ which basically implies that the theory becomes scale-invariant above a certain energy scale  since all couplings converge to a fixed constant value. If a theory demonstrates this behaviour, it is said to be \textit{asymptotically safe}.
		\item \textbf{Freedom}. The second possibility is a special case of asymptotic safety in which the beta functions are negative for all the couplings in the theory and thus the values of the running couplings decrease as the energies become higher and higher until they eventually hit a fixed point which is equal to zero ($g*=0$). Accordingly, for lower energies (and large distances) the coupling strengths increase rapidly until they become divergent at some low, but finite, scale. This behaviour is thought to be responsible for the phenomenon of quark confinement in QCD, and theories of this kind are said to be \textit{asymptotically free}.	
		\item \textbf{Triviality} The third possibility is the case in which the beta function is positive for at least one of the couplings in the theory and there is no non-trivial UV fixed point. This means that the running couplings $g_i(\mu)$ increase indefinitely as the energy scale increases. Theories without a non-trivial UV fixed point are considered \textit{trivial} since their continuum limit is only well-defined if the values of all renormalized couplings, $g_R$, are zero. The physical interpretation of triviality is that the quantum corrections completely suppress the interactions in the absence of a finite cut-off: the physical charges are entirely `screened off' if the ultraviolet cut-off is sent to infinity  \citep{gies:2004,crowther:2022}. A particular problem associated with lack of a non-trivial continuum limit is where the renormalization group flow not only does not hit a fixed point but rather the beta function is found to blow up to infinity at a very large but finite energy level ($\mu_{Landau}$). Such a theory is said to encounter a \textit{Landau pole} \citep{landau:1954}. 
        \item \textbf{Non-Integrability}. A fourth and less explored possibility, is that the RG flow exhibits a cyclic or even chaotic behaviour at higher scales. This possibility was also noted by Wilson himself and although the existence of such cyclic or chaotic RG flows has not been yet confirmed, several models have been developed as candidates for \textit{limit cycle behaviour} \citep{ Curtwright2012,Morozov2003,Leclair2003,Wilson1971}.
        
\end{enumerate}

\cite{Huggett2002} presents a detailed discussion explaining how the renormalization group provides a map that helps us understand the infinities and the `maneuvering' during the renormalization process, making the derived models consequences of the background theory. The basic idea is that one uses the renormalization group and a renormalization scale to generate a family of renormalized theories from a family of bare theories and then confirm that the family of bare theories describes well-defined physics in the continuum limit. The physics of every bare theory is captured by the physics of the corresponding renormalized theory, which is rendered finite by a fixed cut-off limit, and hence the physics of the limit of the bare theories is captured by the limit of the renormalized theories which is finite. The crucial step in the argument is to justify that we have well-defined continuum physics in the limit, and this is achieved, as \citet[p.274]{Huggett2002} explains, by the topology of the parameter space. However, this explanation only holds for quantum field theories that are \textit{asymptotically safe} such as QCD, where the renormalization group flow can be shown to hit a fix point. As Huggett puts it at the very end of his article: `...assuming that a theory is asymptotically safe, the problem is not terminal but can be fixed using renormalization to tune it to a suitable point. If, however, a theory is not asymptotically safe, then [...] renormalization as described here will not be possible' (\textit{ibid.}, p.275-6).\footnote{Here it is worth mentioning the work of \cite{Franklin2020} on the effectiveness of Effective Field Theories. Commenting on the paper of \cite{Butterfield2014b}, Franklin notes that the appeal to the fixed point structure of the RG method in order to justify renormalization is ill suited. His claim is that RG is best seen as a mathematical framework that \textit{codifies} rather than explains the renormalizability of theories in QFT, in that it allows us to mathematically establish whether or not a theory is renormalizable, but nonetheless, it does not provide a sufficient explanation of the physical reasons behind renormalization. Our focus here is on mathematical relationships rather than explanation and in Franklin's terms, our point could be re-expressed as a statement that without a non-trivial UV fixed point the RG method cannot not successfully achieve its codifying role.}

Unfortunately, this rather attractive theoretical diagnostic picture is very difficult to fully implement in practice. For realistic QFTs there is no fully reliable way to explicitly calculate the beta functions to all orders. Take the example of QED. To render the introduced idealizations controllable, one needs to show that the theory is asymptotically safe by calculating the corresponding beta functions in realistic QED models, however, no rigorous proof for the asymptotic safety of QED has been achieved. Rather, what we have available are further mathematical idealization  techniques for approximating the beta functions in some cases, which further \textit{suggest} a possible behaviour of the theory in higher energies. Interestingly, the situation is then that we are attempting to justify the assumptions introduced by regularization and renormalization, by introducing further assumptions, creating a form of regress in which idealizations are justified by further idealizations.\footnote{This is not to say that this is a vicious regress, however. Whether the justification of an approximation by means of a further approximation is epistemically warranted is an interesting open question and deserves to be studied on its own merit in future work. As already noted, the possibility for precisely this situation was anticipated in \cite[p. 293]{hartmann:2001}.} 

In any case, even ignoring the regress worry, the evidence gained from approximate methods for calculating the beta functions is not in fact promising for the case of QED. In particular, calculations at the one-loop level indicate the existence of a Landau pole \cite[p.37-41]{hollowood:2013} and thus provide evidence \textit{against} the asymptotic safety of QED. One might hope, at this point, that the Landau pole is an artefact of the perturbation calculation of the beta function and there is some evidence in this regard \citep{goeckeler:1998}. However, no indication of a non-trivial UV fixed point for QED has been found so far. Furthermore, evidence from non-perturbative lattice simulations \citep{goeckeler:1998,kim:2001} and the exact renormalization group approach \citep{gies:2004} supports the triviality of the theory with complete charge screening if the UV cut-off is sent to infinity. 

The standard response to the presumed lack of asymptotic safety in quantum field theories found in the physics literature is that such theories should not be understood as fundamental theories and a consistent UV completion can be obtained by appeal to more fundamental theories. However, it is \textit{not} expected that electroweak theory can provide a UV completion of QED and, indeed, the triviality problem is expected to also occur in the Yukawa-Higgs sector.\footnote{See for example \cite{hambye:1997} A full list of references is given in \cite{eichhorn:2018} reference [4].} Rather, the hope for a UV completion of QED relies upon appeal to physics beyond the standard model, such as in asymptotically safe quantum gravity scenarios \citep{christiansen:2017,eichhorn:2019}. Whether or not such a response to the QED triviality problem ultimately proves successful is, however, tangential to the issue at hand: we are concerned with the justification for the idealizations involved in the theories and models in the context of quantum field theories, not exogenous justifications of the idealizations coming from other theories, cf. \cite{crowther:2022}. Moreover, since this putative justification would need to come from theories we \textit{do not in fact have}, the supposed resolution has the status of a promissory note. For the case of QED the mathematical idealization in question is evidently uncontrolled (since no fixed point has been identified), and potentially also uncontrollable (since there is evidence that none in fact exists and the theory is trivial). 

Let us return to our principal question: whether the empirically relevant models that are eventually put to test by comparison with experimental data, can still be thought of as theoretical models of the background theory, based on the initial non-perturbative models of QFT. Given that one requires a secure deductive or approximate derivation relation between the background theory and its models, one needs to show that each of the aforementioned introduced assumptions is a controlled idealization. For theories that have been shown to be asymptotically safe to some standard of rigour, there is a strong case that the relevant idealizations are controllable. However, in cases such as QED, where asymptotic safety has not been established or indeed there are indications of triviality, the idealizations involved are uncontrolled. As such, the empirically relevant models of these theories are not theoretical models in the canonical sense.

\section{Implications}\label{implications}

The analysis provided thus far has important implications for the philosophical understanding of theoretical models and the empirical adequacy of scientific theories. Let us recap the argument provided in Section \ref{argument}. The first premise, P1, was that quantum field theory is a more general framework from which specific theories such as quantum electrodynamics and quantum chromodynamics are derived via particular action principles. The second premise, P2, was that a theory has theoretical models if its predicted empirical quantities are derivable either as deductive consequences of the theory or as a result of an approximate derivation involving controllable mathematical idealizations. The third premise, P3, was that we have good reasons to believe that the derivation of predictions from empirically successful theories in contemporary particle physics, such as quantum electrodynamics, involves ineliminable appeal to uncontrolled idealizations. These three premises together lead us to the deeply unintuitive conclusion, C1, that we have good reasons to believe that theories in contemporary particle physics, such as quantum electrodynamics, do not have theoretical models. 

The natural response for those seeking to reject the unintuitive conclusion is to challenge the controversial premise P3. However, as we saw in the previous section, P3 is, as it stands, remarkably resilient, even in the face of the sophisticated framework of renormalization group theory. 

An alternative response would be to appeal to the possible existence of models of QED that are related to theory via as yet unconcieved deductive relations or controlled idealizations. That is, since, logically speaking, whether theoretical models exist or not is independent of our epistemic position, such models may exist even if we are not, and indeed may never be, in a position to understand them. The plausibly of such a response will be heavily dependant on one's background philosophy of science in particular with regard to the ontology of theories and models. If one thinks of theories and their models as real existing abstract objects then this response might allay some fears \citep{giere:1988,psillos:2011}. By contrast, under a quietist or pragmatic stance regarding what it is for models to exist, the response is entirely empty \citep{french:2010,antoniou:2021}. In any case, our conclusion that contemporary particle physics includes theories \textit{which we have good reasons to believe} are theories without theoretical models is unaffected by the speculative possibility that such models may exist even if we have no good reasons to believe that they do. 

It might therefore seem like the only choice left is to `bite the bullet' and accept the conclusion of theories without models, despite its unintuitiveness. However, as we will argue now, this option is undesirable since C1 has further clearly unacceptable consequences for empirical adequacy. Consider the following extension of our argument:

\begin{enumerate}
\item [C1.] Contemporary particle physics includes theories, such as quantum electrodynamics, which we have good reasons to believe are theories without theoretical models.  
\item [P4.] A theory is empirically adequate only if it possesses theoretical models that agree with experimental outcomes.\footnote{cf. \cite[p.12]{vanfraassen:1991}: `To be empirically adequate is to have some model which can accommodate all the phenomena'.}
\item [C2.] We have good reasons to believe that the theory of quantum electrodynamics is not empirically adequate.
\end{enumerate}

Clearly, C2 is not just unintuitive but absurd and thus unacceptable. Quantum electrodynamics is variously understood as the most empirically successful theory humanity has yet constructed, so any argument that ends up with the conclusion that we have good reasons to believe it is not empirical adequate must have gone wrong somewhere! Given that our robust defence of P3 is accepted and that P1 is also accepted as uncontroversial, the two remaining options to avoid absurdity are then to put pressure on P2 and P4. Let us consider each of these options in turn.

If theoretical models are not required to be connected to their background theories via controlled idealizations, then one clearly avoids the absurd conclusion since one is able to understand the empirically relevant models of quantum electrodynamics as models of the theory. The cost to the coherence of our concepts is, however, catastrophic. This is because the expansion of the definition of a theoretical model to include links via uncontrollable mathematical idealizations, \textit{simpliciter}, allows arbitrary mathematical connections between scientific theories and models. To put it differently, saying that the models of a theory include models which are the products of both controllable and uncontrollable idealizations without further qualifications, allows any possible mathematical connection of any possible model with a background theory. Such a view would lead to a narrowly instrumentalist understanding of the empirical adequacy of theories which is solely defined in terms of the empirical success of the derived models, regardless of whether such models are constructed based upon rigorous and well justified scientific and mathematical processes or not. Consequently, under such view, any scientific theory in which various \textit{ad hoc} mathematical modifications are introduced with the sole purpose of achieving agreement with data can be considered as empirically adequate insofar as the derived predictions agree with measurement outcomes. We do not, therefore, take the route via weakening P2 to be worth pursuing. 

A plausible alternative is to distinguish between appropriate and inappropriate uncontrolled idealizations in order to avoid arbitrary mathematical simplifications of models. That is, it may be argued that the models of a theory are only those models in which the introduced mathematical idealizations, although uncontrolled, are still \textit{motivated by theoretical reasons}. This approach succeeds in distinguishing between cases in which mathematical idealizations are completely arbitrary (i.e., unmotivated), explicitly known to be problematic (i.e., we have theoretical reasons \textit{against} them), and cases in which the scientific community has good reasons to believe that the introduced idealizations, although uncontrolled, are plausible (i.e. motivated by by theoretical reasons). In fact, this is indeed precisely what happens in the framework of the renormalization group method where a possible justification for the truncation of the series in sought in terms of asymptotic safety and the framework of effective field theories. 

This idea would lead to a more relaxed definition of empirical adequacy based on the distinction between \textit{appropriate} and \textit{inappropriate} uncontrolled idealizations. Such an approach would require a robust framework for distinguishing between uncontrollable idealizations that are appropriate for the derivation of empirical predictions  and other arbitrary and \textit{ad hoc} idealizations to make models mathematically tractable. Although intuitively some idealizations seem plainly wrong, e.g. the introduction of an arbitrary number of free parameters to fit the data, the presented analysis of the status of idealizations in quantum electrodynamics indicates that there is not as yet a clear basis to determine whether a certain uncontrolled mathematical idealization is appropriate or not in the context of QFTs. The problem is that the concept of `appropriate' and `theoretically motivated' idealization is extremely difficult to define in a rigorous and unambiguous way leading to a clear definition of theoretical models. What counts as appropriate and theoretically motivated depends to a large extend on the subjective appraisals of the involved scientists. We remain sympathetic to this option however, and acknowledge that further work is required to clarify these notions. 

The final option available to us is then to reconsider P4 and provide a more generalised definition of empirical adequacy that avoids the absurd conclusion C2, even if it does not allow us to avoid the unintuitive conclusion C1. The problem however, is that at least within the context of constructive empiricism it is not at all clear how one might reconstruct empirical adequacy. Recall that within the framework of constructive empiricism a theory is empirically adequate if the theoretical models of the theory provide predictions that agree with the contents of measurement outcomes. As \cite{vanFraassen1980} puts it: `[t]o present a theory is to specify a family of structures, its models; and secondly, to specify certain parts of those models (the empirical substructures) as candidates for the direct representation of observable phenomena' (p.64). There appears to us no natural way to generalise empirically adequacy to avoid P4 within the constructive empiricist view of theoretical structure. Of course, whilst constructive empiricism is probably the philosophical view with the most clear and well developed definition of empirical adequacy, there is nothing forbidding us to search for a wider definition.  

One option would be to try and find means by which to connect the empirically relevant models to the theory via reference to further theories or frameworks. That is, to modify the definition of empirical adequacy to allow the models on which the adequacy is based to be linked to the theory via exogenous arguments, which may include idealizations controlled via other theories or frameworks. This approach has the advantage of mirroring the attitude of many practicing scientists and has the interesting broader implication that perhaps philosophers should look beyond the theory as a unit of analysis in the context of evaluations of empirical adequacy. 

In the context of the failure of Borel summability, the exogenous control approach could plausibly be understood to be illustrated by the idea of appealing to non-perturbative `instanton' corrections as a means to control the idealization within the perturbative theory \citep{marino:2014}. The key idea is to use properties of the non-perturbative solutions to understand the failure of Borel summability. In particular, we can differentiate two types of situation when there is no Borel summability: i) we are considering perturbative series around \textit{unstable minima}; and ii) there are \textit{extra saddle points} in the path integral. The second situation is understood to correspond to non-perturbative effects of the instanton type. In that context, one can consider `lateral Borel resummations' (i.e. integrals along paths in the complex plane) and the ground state energy can be reconstructed from the Borel-resummed perturbative series corrected by Borel-resummed instanton solutions \cite[p. 37]{marino:2014}. The idea is, then, that realistic perturbative quantum field theories which are \textit{not} Borel summable, admit a non-perturbative analysis with instanton corrections (or generalisations thereof).  Formally, the relevant properties have only been shown to hold in toy models and thus the idealization in question is certainly not controlled as yet. However, there may be scope for experimental demonstration of the existence of instantons in the standard model \citep{amoroso:2021,tasevsky:2023} and this would go a long way towards provision of the kind of control that we take the idealizations of perturbative quantum field theory to require in order to anchor their empirical adequacy. Furthermore, one might plausibly argue that instantons are \textit{not really exogenous}: even though they amount to novel particle content and involve going beyond the perturbative framework, instantons are still part of the framework of QFT in general terms. As such, one might argue they constitute a means for endogenous control. This issue is also worthy of further analysis. 

There are also, as already mentioned, plausible possibilities for exogenous arguments that could provide control of the idealizations in the context of renormalization. These also come from a non-perturbative analysis but are exogenous to an even greater degree since they require us to include additional effects due to gravity. In particular, in the context of failure of asymptotic safety in quantum electrodynamics, scientists in the field of quantum gravity propose to control of the relevant idealization by appeal to the existence of a fixed point in a unified theory of quantum gravity. This wider definition would be aligned to the general attitude that UV divergences of QFT point to quantum gravity as discussed by \cite{crowther:2022} and to 
the particular motivations for the asymptotic safety programme \citep{eichhorn:2019}.\footnote{See also \cite{crowther:2019}.} 

Whichever option is pursued, we take ourselves to have demonstrated that the problems for the articulation of theoretical models and empirical adequacy in the context of perturbative quantum field theory are severe ones and warrant serious attention from philosophers and, perhaps, also scientists themselves.   

\setcounter{tocdepth}{-1}
\section*{Acknowledgements}
	
We are greatly appreciative to Richard Dawid, Sean Gryb and Tim Koslowski for extremely helpful comments on a draft manuscript. Earlier versions of this work have also greatly benefited from valuable feedback from James Fraser, Michael Miller and Alex Franklin to whom we express our gratitude. The paper also greatly profited from feedback from three anonymous reviewers.

\subsubsection*{Conflict of Interest Statement}

There are no conflicting or competing interests.

\bibliography{ModelsQFT_1}
\bibliographystyle{chicago}

\end{document}